\documentclass[preprint]{aastex}
\usepackage{graphicx, natbib,epsfig,amsmath,amssymb, mathrsfs, txfonts}
\usepackage{epstopdf}
\begin{document}

\title{Band-Limited Coronagraphs using a Halftone-dot process:\\ design guidelines, manufacturing, and laboratory results}

\author{P. Martinez\altaffilmark{1}, C. Dorrer\altaffilmark{2}, and M. Kasper\altaffilmark{1}}

\altaffiltext{1}{European Southern Observatory, Karl-Schwarzschild-Strasse 2, D-85748, Garching, Germany}
\altaffiltext{2}{Aktiwave, 241 Ashley drive, Rochester, NY, 14620-USA}

\begin{abstract}
The Exo-Planet Imaging Camera and Spectrograph (EPICS) for the future 42-meter European-Extremely Large Telescope, will enable direct images, and spectra for both young and old Jupiter-mass planets in the infrared.
To achieve the required contrast, several coronagraphic concepts -- to remove starlight -- are under investigation: conventional pupil apodization (CPA), apodized-pupil Lyot coronagraph (APLC), dual-zone coronagraph (DZC), four-quadrants phase mask (FQPM), multi-stages FQPM, annular groove phase mask (AGPM), high order optical vortex (OVC), and band-limited coronagraph (BLC). Recent experiment demonstrated the interest of an halftone-dot process -- namely microdots technique -- to generate the adequate transmission profile of pupil apodizers for CPA, APLC, and DZC concepts. Here, we examine the use of this technique to produce band-limited focal plane masks, and present guidelines for the design. Additionally, we present the first near-IR laboratory results with BLCs that confirm the microdots approach as a suitable technique for ground-based observations. 
\end{abstract}


\section{Introduction}

By the end of 2018, challenging projects such as EPICS (Exo-Planet Imaging Camera and Spectrograph, \citet{EPICS}) for the future 42-meter European-Extremely Large Telescope, or PFI  \citep[Planet Formation Imager,][]{M06} for the Thirty Meter Telescope (TMT) will enable direct images, and spectra for warm self-luminous and reflected-light Jovian planets. These instruments will operate with eXtreme Adaptive Optics system (XAO) designed for high Strehl ratios (i.e. $\sim90\%$ in $H$-band), as for SPHERE \citep{2008SPIE.7014E..41B} and GPI \citep{2006SPIE.6272E..18M}, forthcoming planet-finder instruments with first light planned in 2011. 

Several coronagraph concepts have been studied extensively in the past years, with the objective of finding optimized designs that can sufficiently suppress the on-axis starlight, allowing faint companions direct detection (e.g. \citet{Malbet96, SIVA01, Guyon07, Corono}). 
Among them, the band-limited coronagraph \citep[][BLC]{2002ApJ...570..900K} has been proposed to completely remove starlight. The BLC has the advantage of being less sensitive to the primary mirror segmentation, unavoidable with ELTs, than for other concepts \citep{SIVA05}, and to provide achromatic behavior. 
Additionally, in \citet{Corono}, we pointed out the interest of such concept on an ELT for either very bright object detection, or for the search of planets at large angular separations.
 
Several BLC prototypes have been developed during the past years \citep[e.g.][]{Debes04b, Trauger, Creep06, Trauger2, Moody08} for visible wavelength application.
Several technical approaches have been used: (1/) gray-scale pattern written with an high-energy beam sensitive glass (HEBS) using e-beam lithography, (2/) Notch filter pattern -- binary mask -- written with thick Chromium layer on a substrate, dry-etched with high density decoupled plasma, (3/) gray-scale pattern manufactured with vacuum deposited metals and dielectrics.  
Even if electron-sensitized HEBS glass can accurately accommodate continuous range of transmission, the darkening of the HEBS glass under electron bombardment is accompanied by a determined phase shift, while the technique suffers from a lack of experience in the near-IR. 
Same constraints apply for the vacuum deposited metal technique.
The notch filter has the advantage of being intrinsically achromatic.
These designs, consisting of a particular implementation of small structures (stripes of opaque material with width of about some microns) must be finely controlled in size, spacing, and opacity.
Mask errors and tolerance are discussed in \citet{Kuchner03}, where requirements might be strong for near-IR application. 

In this paper, we examine the use of a halftone-dot process, namely microdot technique, to reproduce a continuous mask profile as already done for pupil apodizer for SPHERE \citep{microdots1, microdots2}, and being manufactured for the JWST NIRCam coronagraph \citep{Krist09}.
These masks consist of distributions of opaque square pixels (called dots) to reproduce the continuous transmission of a filter with several advantages:
relative ease of manufacture, achromaticity, reproducibility, and ability to generate continuous transmission ranges, without introducing wavefront errors. Besides, mask errors can be easily pre-compensated \citep{2007JOSAB..24.1268D}.  

Section 2 describes the microdot design principle and properties, while Sect. 3 provides guidelines for the design. Section 4 presents monochromatic, and polychromatic results obtained in laboratory in the near-infrared.
Finally Sect. 5 concludes on the suitability of the microdots approach for producing BLCs in the context of ground-based instruments. 

\section{Principle}
\label{principle}
In its general scheme, a microdots filter is an array of dots (i.e. pixels) that are either opaque or transparent. It is fabricated by lithography of a light-blocking metal layer deposited on a transparent glass substrate. 
To best approximate specifications, the dot distribution is regulated by a Floyd-Steinberg dithering algorithm \citep{floyd}, based on error diffusion principle. This allows the accurate generation of grey levels, and rapidly varying shaping functions.  
This algorithm is presented, and discussed in \citep{Ulichney, Ulichney2} for printing technique applications, as well as for laser beam shaping \citep{2007JOSAB..24.1268D}. Additionally, this algorithm was successfully used for producing microdots pupil apodizers for the apodized pupil Lyot coronagraph \citep{microdots1, microdots2}. 

\begin{figure*}[!ht]
\begin{center}
\includegraphics[width=14cm]{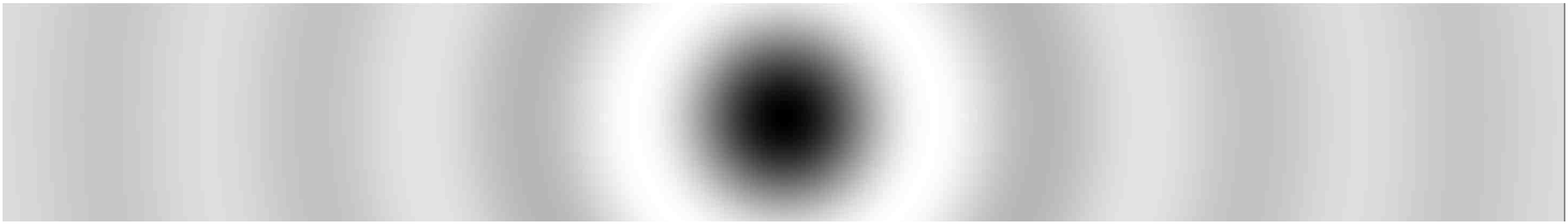}
\includegraphics[width=14cm]{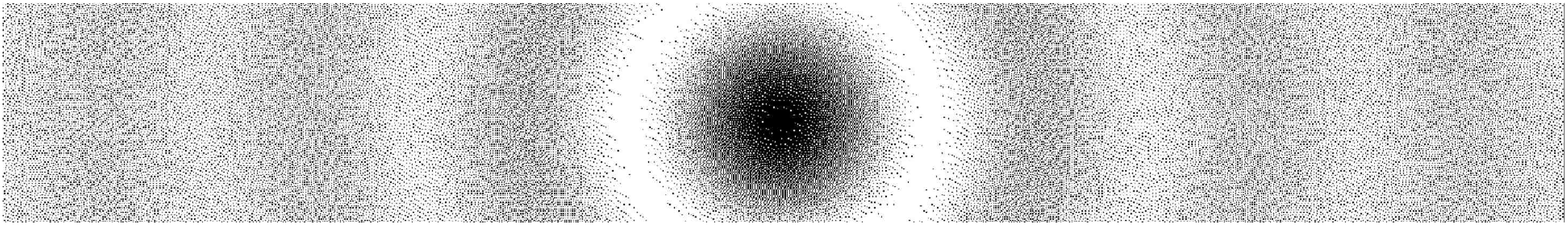}
\end{center}
\caption{Simulation maps of a circular band-limited mask. $Top$: Scan of the center (grey level pattern). $Bottom$: corresponding optimized microdots pattern. Grey-scale color evolves from $0\%$ (black) to $100\%$ (white) transmission.} 
\label{view}
\end{figure*}

By principle, the microdot mask -- the BLC amplitude function -- is sampled, subject to the dot size (see Fig. \ref{view}). This issue will be further discussed in Sect. \ref{design} from a design specification point of view.
In the following, we will examine its impact on the coronagraphic effect by comparing to that of a continuous transmission mask.

We consider a telescope with aperture function $A$, and a band-limited amplitude mask function from  \citet{2002ApJ...570..900K}:
\begin{equation}
M(r) = N \left( 1 - sinc\left(\frac{\epsilon r D}{\lambda}\right) \right)
\label{function}
\end{equation}

\noindent where $\lambda$ is the wavelength of the application, $r$ the radial coordinates in the image plane, D the telescope primary diameter, $\epsilon$ the bandwidth which rules the inner-working angle of the coronagraph (IWA hereafter), and finally N is a constant of normalization insuring that $0 \leqq M(r) \leqq 1$.

\noindent The Fourier transform of $M(r)$ denoted $\mathscr{M}(u)$, is therefore:
\begin{equation}
\mathscr{M}(u) = N \left( \delta \left( \frac{u \lambda}{D}\right) - \frac{ \lambda}{\epsilon D} \times \Pi \left(\frac{u \lambda}{\epsilon D} \right)  \right)
\label{function}
\end{equation}
\noindent where $u$ states for the radial coordinate of the pupil plane, $\delta$ the Dirac-function, and $\Pi$ the top-hat function.

\begin{figure*}[!ht]
\begin{center}
\includegraphics[width=8cm]{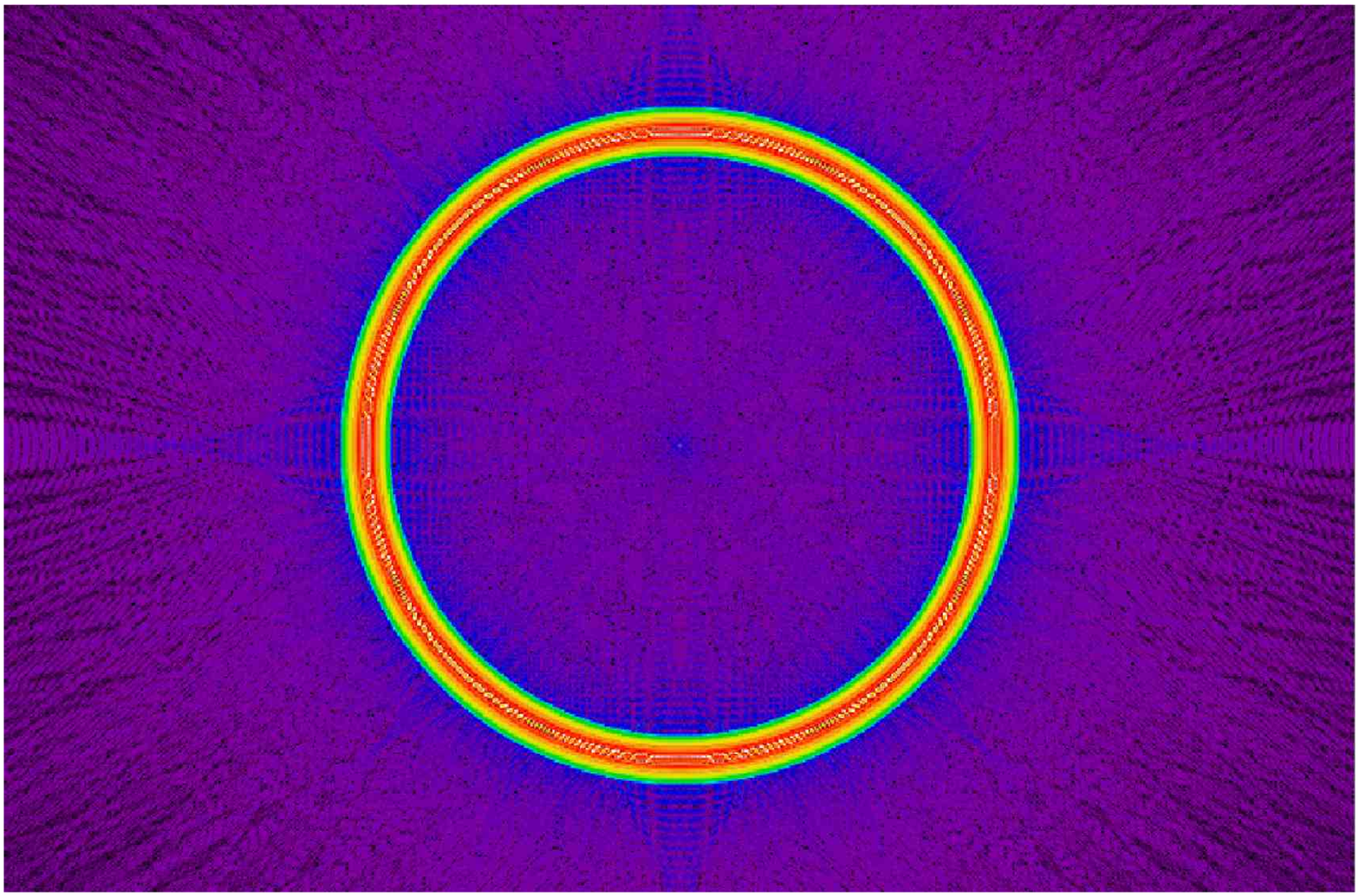}
\includegraphics[width=8cm]{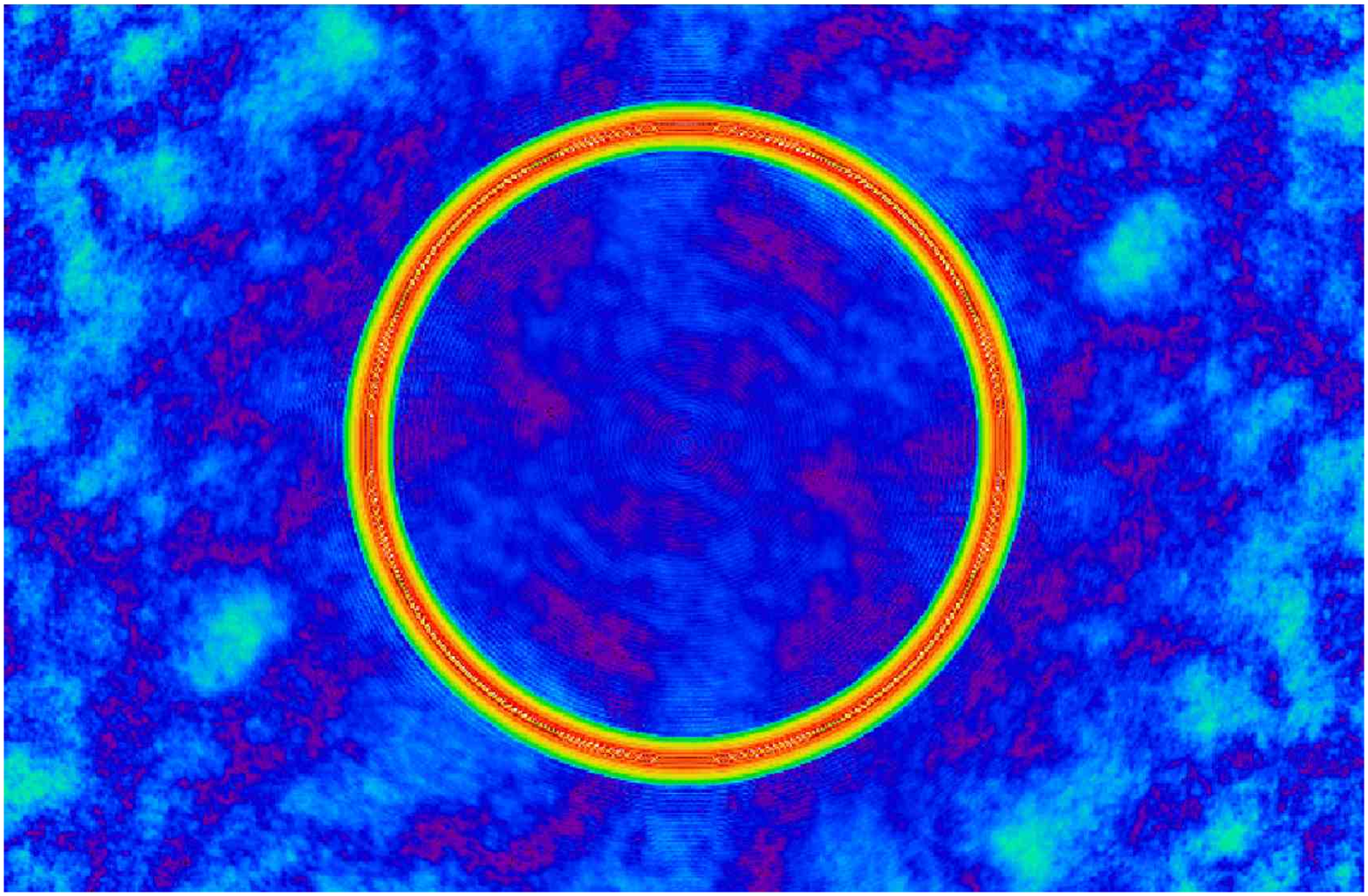}
\includegraphics[width=8cm]{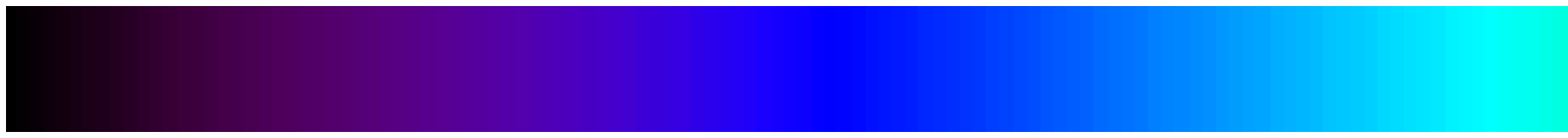}
\includegraphics[width=8cm]{./FIGURES/scale2.eps}
\end{center}
\caption{Example of pupil plane images before effect of the pupil-stop, with a continuous BLC (left), and a microdot BLC (right). The arbitrary false color distribution has been chosen to enhance contrast for the sake of clarity. The scale evolves from noise-level (black / pink colors) to maximum intensity (red-color).} 
\label{pupilimages}
\end{figure*}

\noindent In a BLC, the incident electromagnetic field propagates from the telescope aperture to a focal plane where the mask, $M(r)$, is applied. After the occulting mask (filtering for the low-frequencies), the pupil is re-imaged in a second pupil plane where a pupil-stop is placed (high-frequency filter). Then the scientific image is recorded in a second focal plane where the detector is installed.
The coronagraphic effect can be well described in the second pupil plane ($\psi_{pp}$, hereafter), as the convolution product of $\mathscr{M}(u)$ by the telescope aperture $A(u)$, times the pupil-stop function denoted $S(u)$, such as:
\begin{equation}
\psi_{pp} =  [ \mathscr{M}(u) \star A(u) ] \times S(u)
\label{Eqpupil}
\end{equation}
\noindent where the star ($\star$) states for the convolution product.
The power of band-limited coronagraph comes from the properties that $\mathscr{M}(u)$ is equal to zero everywhere but $\vert u \vert < \frac{\epsilon D}{2 \lambda}$, i.e. the power spectrum of $M(r)$ has power in a limited domain of frequencies. As a result, in the second pupil plane (Eq. \ref{Eqpupil}), the convolution product of $\mathscr{M}(u)$ by the pupil aperture confines the diffracted light in the vinicity of the pupil edges as exhibits in Fig. \ref{pupilimages} (left, where the pink color states for numerical noise level). This energy can be completely removed with the pupil-stop, $S(u)$, similar to the aperture function, but reduced in diameter. 
A development of the BLC theory can be found in \citet{2002ApJ...570..900K}, although we note that the present formalism assumes that the amplitude mask is purely real, i.e. it does not take into account the relationship between opacity and phase shift in real physical material, which impacts the performances.

\noindent A microdot BLC will have a different power spectrum than a continuous mask because its function is sampled by the dots, which forces its Fourier transform to be periodic, and the algorithm used to distribute dots introduce high-frequency noise. As a result, the convolution product of $\mathscr{M}(u)$ by the pupil aperture still confines the diffracted light in the vinicity of the pupil edges, but some energy remain inside and outside the geometrical pupil as exhibits in Fig. \ref{pupilimages} (right).
The residual part inside the geometrical pupil, not filtered by the pupil-stop, will therefore set a limit in the achievable performance. The intensity of the residual energy as seen in Fig. \ref{pupilimages} (right) is 4 to 5 order of magnitude bellow the maximum intensity of the ring surrounding the pupil aperture.
The amount of residual energy in the pupil plane is a function of the dots size, and the bandwidth of the function ($\epsilon$), since these two parameters directly impact the sampling of the design. This point will be further analyzed in Sect. \ref{dotsize} with simulations.

\noindent Although BLC designed with microdots can not achieve perfect cancellation of starlight, it is important to note that a ground-based instrument does not require ideal performance of a BLC, but can accomodate with relaxed performance since limitation will first come from the quasi-static speckle halo level. Planet-finder instruments will use speckle calibration strategies to improve contrast delivered after correction of the turbulence by an Adaptive Optic system (AO).
If a simultaneous speckle rejection technique is used (e.g. simultaneous differential imaging, spectral deconvolution...), as selected for SPHERE \citep{2008SPIE.7014E..41B}, GPI \citep{2006SPIE.6272E..18M}, and EPICS, 
performance of a coronagraph does not necessarily need to be better than this quasi-static speckle halo level ($10^{-5}$ to $10^{-6}$). A deeper contrast is then achievable through appropriate data reduction.

\section{Mask design guidelines}
\label{design}
\subsection{Specified functions}
\begin{figure}[]
\begin{center}
\includegraphics[width=10cm]{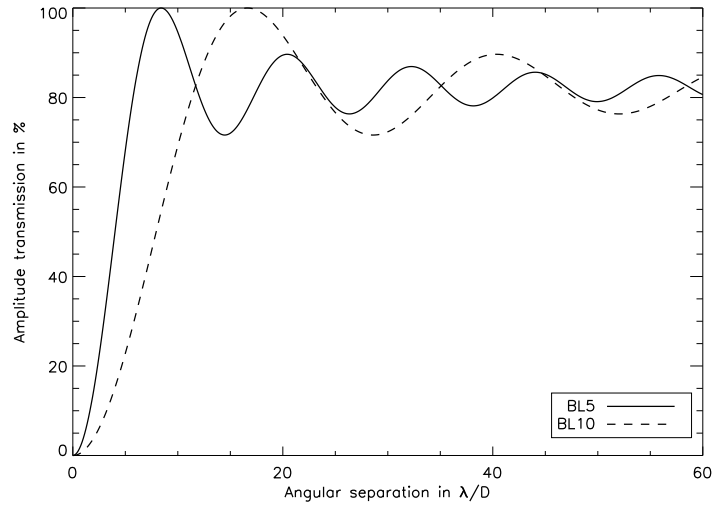}
\end{center}
\caption{Band-limited functions used in simulation, and in the experiment.} 
\label{functions}
\end{figure}

We consider two BLC masks with identical function (Eq. \ref{function}), but different bandwidth (i.e. different $\epsilon$ values, see Fig. \ref{functions}).
Hereafter, BL5 and BL10 correspond respectively to $\epsilon=0.17$ (IWA = 5$\lambda/D$) and $\epsilon=0.086$ (IWA = 10$\lambda/D$). Even if BL10 has a large IWA ($\sim 0.08$ arcsec in H-band for a 42-meter telescope), only suitable to the search of companions at large angular separations, it has a smoother transition in between the low and high-transmission parts of its profile, and can be compared with BL5 in the light of the technique employed (i.e. sampling issue).

\subsection{Size of the dots}
\label{dotsize}
\begin{figure*}[!ht]
\begin{center}
\includegraphics[width=10cm]{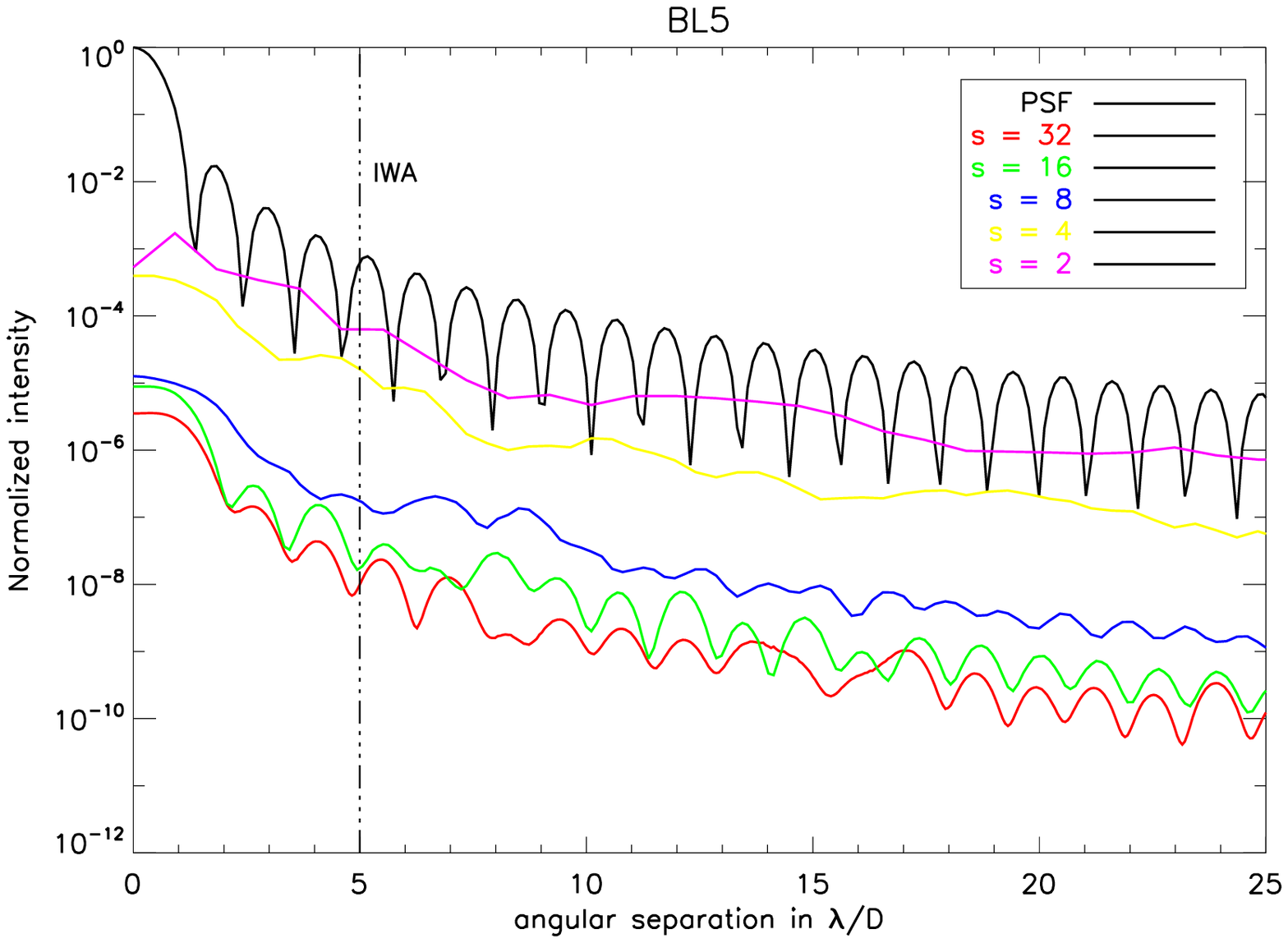}
\includegraphics[width=10cm]{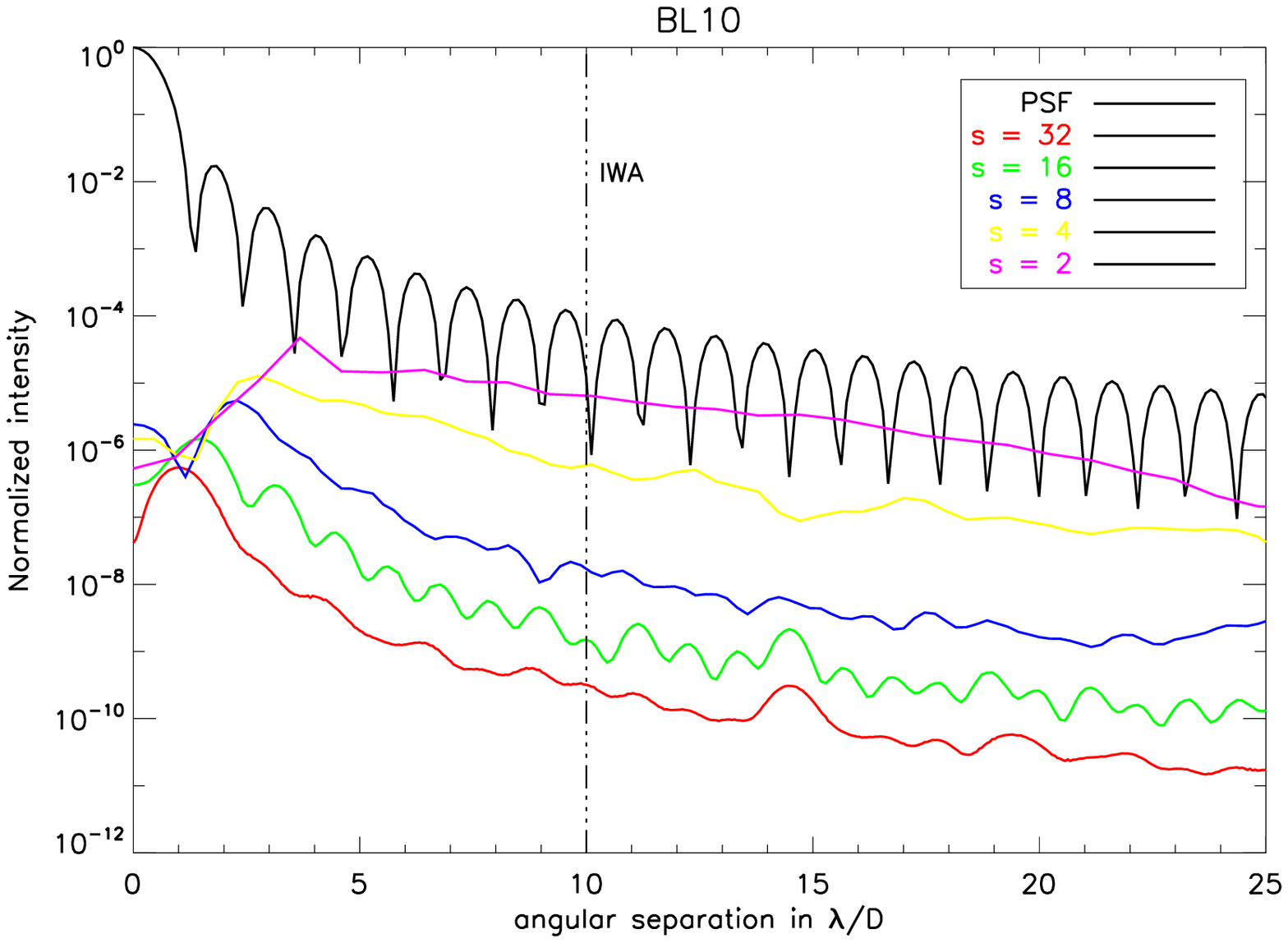}
\end{center}
\caption{Impact of the dot size on the coronagraphic performance for BL5 (top) and BL10 (bottom).} 
\label{simul}
\end{figure*}

The selection of the dots size is in first approximation -- as discussed in Sect. \ref{principle} -- formally equivalent to a sampling problem. 
Better results are obtained with smaller pixels, since this allows finer control of the local transmission.
This is particularly an important issue in region of the mask where transmission is very low (center of the mask). 
Since BLC masks are placed in the focal plane of an instrument, their dimensions are usually expressed in $\lambda/D$ units.
Therefore, a simple metric to quantify the sampling of the function is the ratio of the Airy unit by the dot size ($p$).
Optimally, this metrics must be defined at the shortest wavelength for which the mask is designed to operate, and can be expressed as:
\begin{equation}
s = \frac{F_{\#} \times \lambda_{min}}{p}
\label{E2}
\end{equation}

\noindent where $F_{\#}$ is the $f$-number on the mask and $\lambda_{min}$ the shortest wavelength of the application (i.e. insuring that $s$ remains greater than the specification for all wavelengths). 
Using simulation maps of microdots BLC (assuming specified dot spatial distribution), we analyzed how the dot size affects the coronagraphic performance with respect to continuous idealistic masks.
 Our simulations make the use of Fraunhofer propagators between pupil and image planes, which is implemented as fast Fourier transforms (FFTs) generated with an IDL code.  
Dots are sampled by one pixels to prevent from excessive size of array critical in the high $s$ values domain. Validity of the dot sampling has been verified with simulations assuming different sampling (1, 4 and 16 pixels per dot), as discussed in \citet{microdots1}. An effect is only expected in the very low $s$ domain at high-spatial frequencies, out of the field of view of interest.
These monochromatic, and aberration-free simulations were addressed for the two band-limited configurations we manufactured (BL5 and BL10).
Results are presented in Fig. \ref{simul}. Note that a continuous mask provides a total on-axis source suppression (hence not plotted in the figure) assuming the idealistic conditions of these simulations.
As expected (Sect. \ref{principle}), ideal performance of BLC can not be met (i.e. contrast deeper than $10^{-12}$, in Fig. \ref{simul}). 
However, several $s$ configurations (8, 16, and 32) yield interesting contrast levels, deep enough to meet contrast requirements of planet-finder projects.
Considering our application (H-band, $\lambda_{0}$= 1.6 microns, with a contrast requirements of $10^{-6}$ at 5$\lambda/D$ after the coronagraph, e.g. EPICS), suitable configurations are $s=8$ and $s=16$, as the $s=32$ situation yields to very small dots size (2.5 microns) comparable to the size of the wavelength of light. In this particular situation, it is difficult to predict how the field will react to such grating, where scalar diffraction theory does not apply.
Additionally, since $\epsilon$ impacts the gradient from the very low-transmission maintained at the center to the high-transmission part (bandwidth of the function), better performance are obtained when $\epsilon$ is larger (Fig. \ref{simul}, left and right). 
The issue is that the very low transmission maintained at the center of the BLC functions is significantly impacted by the dot size.  
This point will be further discussed in Sect. \ref{epsilon}.

\subsection{Function bandwidth}
\label{epsilon}
\begin{figure}[]
\begin{center}
\includegraphics[width=10cm]{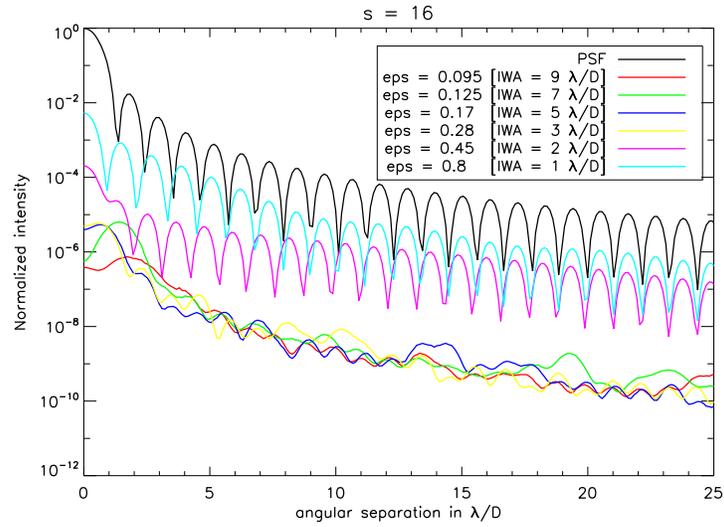}
\end{center}
\caption{Impact of the mask bandwidth on the IWA (left) and performance (right).} 
\label{simul2}
\end{figure}

The bandwidth parameter of BL functions ($\epsilon$) directly impacts the gradient from the low to the high-transmission part of the profile, and rule the IWA of the coronagraph, it is therefore important to analyze its effect on the performance. We simulated several cases with different $\epsilon$ values (from 0.095 to 0.8, i.e. IWA from 1 to 9 $\lambda/D$, see Fig. \ref{simul2}) for the $s=16$ case.
Results gathered in Fig. \ref{simul2} show that for IWA larger than 3$\lambda/D$ ($\epsilon<0.28$) performance in the halo are not $\epsilon$-dependent, while only a slight difference is observable on the peak. For IWA in between 3 and 2$\lambda/D$, although an important degradation of performance appears, coronagraphic capabilities are not negligible (contrast of $10^{-5}$ at IWA). Under $2\lambda/D$ IWA, situation degrades further.

\subsection{Dot opacity}
Whatever the regime of wavelength for which the device is designed, the optical density (OD) is a critical issue to avoid leaks of the starlight in the center of the detector image.
The OD is guaranteed by sufficient thickness of the metal layer (denoted $e$ hereafter). For visible application, Chrome layer are commonly used, while for near-IR applications, more opaque material are mandatory (Aluminum, for instance).
The OD is defined at a specific wavelength, and for a given material, and thickness such as:
\begin{equation}
OD(e, \lambda) = \frac{4\pi k(\lambda) e}{\lambda \times ln(10)}
\label{OD}
\end{equation}
\noindent where $k$ describes the linear attenuation of the optical wave (material-dependent), $e$ stands for the propagation distance (i.e. the material thickness), and $\lambda$ is the operating wavelength.
Equation \ref{OD} is further detailed in appendix \ref{appendix}.
The OD must be therefore carefully defined, accordingly with expected performance of the coronagraph ($s$ and $\epsilon$-dependent, Fig.\ref{simul} and \ref{simul2}).

\section{Experiment}
\subsection{Optical setup}
\begin{figure*}[!ht]
\begin{center}
\includegraphics[width=17cm]{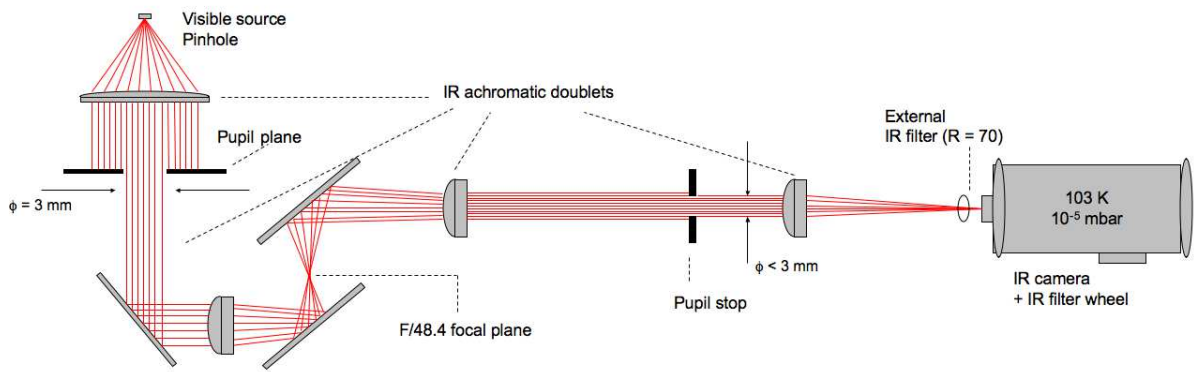}
\end{center}
\caption{Schematic setup of the coronagraphic test-bench.} 
\label{bench}
\end{figure*}

The optical setup is a near-infrared coronagraphic transmission bench developed at ESO (Fig. \ref{bench}). 
All the optics are set on a table with air suspension in a dark room and are fully covered with protection panels forming a nearly closed box.
The entrance aperture is full-filled, and has a 3 mm diameter ($\Phi$) made in a laser-cut,
stainless-steel sheet to an accuracy of 0.002 mm. BLC masks were installed at an F/48.4 beam. Re-imaging optics were made with IR achromatic doublets.
The pupil-stop has a diameter $D_{stop} = 0.84 \times \Phi$, and remains the same during the experiment. Its optimization gave emphasis to the achromatic behavior of BLCs. 
The quality of the collimation in pupil planes was checked and adjusted using an HASO 64 Shack-Hartmann sensor. A pupil-imager system was implemented to align the pupil-stop mask with the entrance-pupil mask to complete alignment in the x- and y- directions. The coronagraphic focal plane was localized using a visible CCD mini-camera with a HeNe laser light and tuned in the final IR image on the detector.
The IR camera used (the Infrared Test Camera) uses a HAWAII $1k\times1k$ detector, cooled to 105 K degree with a vacuum pressure of $10^{-5}$ mbar. Internal optics were designed to 
reach a pixel scale of 5.3 mas (almost 8 pixels per $\lambda/D$). Experiment was done in H-band using either a narrow filter ($\Delta \lambda/\lambda =1.4\%$), or a broadband filter ($\Delta \lambda/\lambda = 24\%$). 
The Strehl ratio of the bench was evaluated to $\sim92\%$. 
It was determined by measuring the peaks intensity ratio of the experimental PSF to that of the theoretical PSF normalized to the total intensity.
The theoretical PSF is created through two different methods, both converging to the same Strehl ratio:
(1/) by performing the foward fast Fourier transform (FFT) of the autocorrelation of an oversampled and uniformly illuminated entrance pupil image from our telescope pupil mask, 
(2/) by performing the FFT of a simulated aperture function with radius determined from the experimental PSF on the basis of photometric criteria.

\subsection{Prototypes}
\begin{figure}[]
\begin{center}
\includegraphics[width=16cm]{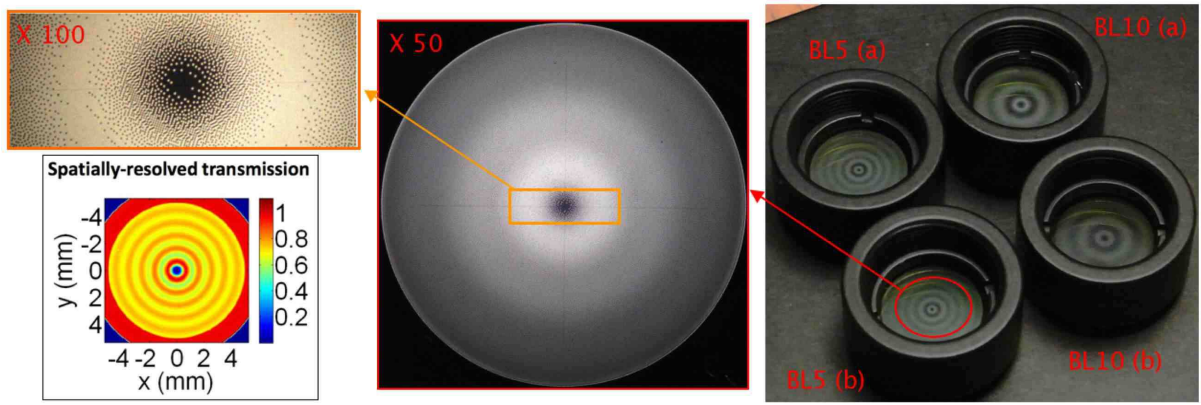} 
\end{center}
\caption{The four prototypes manufactured as seen in their integration mounts (right), successive Shadowgraph inspections (middle and top-left, $\times$ 50 and $\times$ 100), as well as a spatially resolved transmission of BL5 (b) recorded at 1.0 microns (bottom-left).} 
\label{inspection}
\end{figure}

Four prototypes were manufactured by Precision Optical imaging (Rochester, New-York), two BL5, and two BL10 masks. All the masks were designed for 1.64$\mu$m (H-band), and fabricated using wet-etch contact lithography of an Aluminum layer ($OD=8+$, $e=2000\dot{\mathrm{A}}$) deposited on a BK7 substrate ($\lambda/10$ peak-to-valley, 0.5 inch diameter). Antireflection coating has been applied for each faces ($R<0.5\%$ from 1.2 to 1.8$\mu$m).
For each design (BL5, and BL10), two dot sizes have been produced: 5 and 10 microns (i.e., s = 16, and s = 8) denoted by (a) and (b) respectively (see Fig. \ref{inspection}). The OD has been specified to guarantee against leakage greater than the intrinsic limitation of the microdot technique (for s = 16,  and s = 8, see Fig. \ref{simul}).
All the masks have been inspected and cleaned up before integration (Fig. \ref{inspection}).
Profiles were measured at 1.0$\mu$m. The spatially resolved transmission has been obtained after background subtraction and flat field normalization.  
Profile accuracy is of about $5\%$ of the specification. 
The error is mostly localized in the outer part of the mask (high-transmission part), while the center part (for the low-transmissions) is highly accurate (Fig. \ref{profile}). 
The error in the outer part took its origin in a calibration issue of the process, that will be corrected for with new prototypes.

\subsection{Results}
\label{results}
The following metrics are used to evaluate coronagraphic abilities to suppress the on-axis starlight:
\begin{itemize}
\item \textbf{$\tau$} stands for the \textit{total rejection rate}: ratio of total intensity of the direct image to that of the coronagraphic image. 
\item \textbf{$\tau_{0}$} stands for the \textit{peak attenuation rate}: ratio of the maximum of the direct image to that of the coronagraphic image.
\item $\mathscr{C}$ stands for the \textit{contrast}: ratio of the coronagraphic image at a given angular separation to that of the maximum of the direct image to the intensity, azimuthally averaged.            
\end{itemize}

\subsubsection{Data acquisition and reduction}
To achieve high-dynamic range measurements, a serie of 3s short exposure images averaged over 3 mn, and the presence or absence of neutral density filters, were employed. 
Neutral density filters were only applied on non-coronagraphic images. PSFs and coronagraphic images were dark subtracted by turning off the light. 
Data reduction process correct for bad pixels, background, and scales images by the exposure time and optical density.

\subsubsection{Results with BL5}
BL5 prototypes have been tested with the narrow ($\Delta\lambda/\lambda = 1.4\%$), and broadband ($\Delta\lambda/\lambda = 24\%$) filters.
Coronagraphic images are presented in Fig. \ref{images}, while profiles of BL5(a) and BL5(b) are gathered in Fig. \ref{banc} (top).
All coronagraphic runs made with BL5(a) and BL5(b), whatever the filter, roughly yields to identical performance.
BL5 demonstrate an achromatic behavior as expected for such coronagraph.
No difference in between BL5(a) and BL5(b) has been observed, therefore regarding to the performance obtained, $s = 16$ and $s = 8$ are suitable configurations.
A discrepancy of $\sim$1 order of magnitude for both BL5(a) and BL5(b) has been revealed on the peak, while in the halo it evolves between 2 and 3 order of magnitude (see theoretical results as a function of $s$ in Fig. \ref{simul}, left). 
Despite this discrepancy these first results (gathered in Table \ref{resum}) are better than the SPHERE prototypes performance \citep{Reynard}, and experimental results reached on the same bench with an APLC \citep{microdots1}. However, we note that these prototypes (Four Quadrant Phase Mask, and APLC) were tuned for very small IWA (1, and 2.3 $\lambda/D$ respectively compared to 5 and 10 for the BLC described here). Contrast evolves from $\sim3\times10^{-5}$ at IWA to $\sim3\times10^{-8}$ at $20\lambda/D$ (cut-off frequency of SPHERE-like AO system).

\subsubsection{Results with BL10}
Results presented with BL10 prototypes (Fig. \ref{banc}, bottom) correspond to $\Delta\lambda/\lambda = 24\%$. 
As for BL5, we did not notice any difference between performance reached either by BL10(a), or BL10(b). Contrast evolves from $\sim1\times10^{-7}$ at IWA to $\sim1\times10^{-8}$ at $20\lambda/D$, while the peak attenuation rate is about $10^{5}$.    
A discrepancy of about one order of magnitude has been revealed on the peak as for BL5, while in the halo it evolves between 1 and 2 order of magnitude at IWA, and at $25\lambda/D$ depending on which configuration of BL10 we are looking at (Fig. \ref{simul}, left). As expected BL10 results went beyond BL5 performance, although the IWA is quite large. 
The coronagraphic image exhibits speckle halo (Fig. \ref{images}) induced by the non-perfect nature of optics used in the bench.
This point is further discussed in the following section (Sect. \ref{wavefront}).

\subsection{Performance limitations}

\label{discrepancy}
The discrepancy revealed in the experiment may find its origin in several error sources, among them -- for the most important ones --, the scatter-light due to imperfect optical components that create speckles in the scientific image, and mask profile errors.
Note that the source diameter (0.082$\lambda/D$), and the alignment errors of the masks are negligible considering the IWA of the BLCs, while the impact of the filter bandpass is not considered because of the achromatic behavior of BLCs (confirmed in the experiment). We also neglect the fact that the pupil-stop is not black-coated (e.g. black anodization) which may produce potential reflections.
 
\subsubsection{Mask profile errors}
Simulations using real measured profiles (i.e. global profile measured after free-space propagation of the BLC masks and using an imaging optical system, Fig. \ref{profile} for instance), showed that profile errors are actually responsible of one order of magnitude loss on the peak and while the halo is less affected. Assuming their respective measured profile errors, the rejection factor for BL5 and BL10 ($s=16$) are respectively limited to: $\tau_{mask}$ = 6400, and $\tau_{mask}$ = 108000. 

Ideally, going to very small pixel size improves the accuracy of the mask profile transmission (i.e. the sampling problem). 
In practice, two constraints set a limit on the interest of a very small pixel size:
(1/) getting good accuracy becomes more difficult because fabrication errors become more important as the pixel size decreases, 
(2/) when the pixel size is comparable to the wavelength of light, the transmission is affected by plasmons \citep{RCWA1, RCWA2}. In such situation, the transmission might be strongly dependent on the wavelength.
Therefore, it is very likely that efforts must be focused on reducing fabrication errors to avoid an increase of transmission, as a result of a reduction of the metal dots during the isotropic wet-ech lithography process. 
An improvement of the pre-compensation algorithm applied on the digital design, correcting for edge effects on the light-blocking metal dots, by estimating the feature size that would be obtained after fabrication is thus critical.

\subsubsection{Wavefront errors}
\label{wavefront}
The wavefront error (wavefront high-frequency components) of the optical components prior to the pupil-stop imposes an important limitation when reaching high-contrast.
The optical components on the bench upstream of the pupil-stop have a standard quality: flats mirrors of $\lambda/10$ ptv (i.e $\lambda/35$ rms), and achromatic doublets of $\lambda/5$ rms, over the full diameter (25.4 mm) at 633 nm, which corresponds respectively to $\sim \lambda/300$ and $\sim \lambda/45$ rms for the 3 mm pupil diameter (assuming the quality scales linearly with the beam size). Taking into account all components, and assuming that phase added quadratically, it corresponds to an overall approximated total amount of wavefront error of $\mathscr{w} \sim \lambda/67$ rms at 1.6 microns, (i.e. 24 nm rms). With simulations using a Fraunhofer propagator IDL code, and assuming theoretical mask profile, and using typical power-law of manufactured optics, described by PSDs with $f ^{-2}$ variation \citep{2002ApOpt..41..154D}, where $f$ is the spatial frequency, we actually found that 12 nm rms prior to the pupil-stop, already imposes a contrast floor in between $10^{-7}$ and $10^{-8}$, which is in good agreement with the wavefront error estimation discussed above, which is commonly used as a rule of thumb for order of magnitude estimation, as performed in \citet{2003PASP..115..712R}. 
For 12 nm rms, the total rejection factor for BL5 and BL10 ($s=16$) are respectively limited to: $\tau_{wavefront}$  = 40400, and $\tau_{wavefront}$ = 188200.

\subsubsection{Performance estimation vs. experimental results}
\noindent Assuming that these independent errors are added quadratically, the global rejection factor of BLs including mask error, and wavefront errors is:
\begin{equation}
\tau = \frac{1}{\sqrt{ \left( \frac{1}{\tau_{mask}^{2}} +  \frac{1}{\tau_{wavefront}^{2}} \right)}}
\end{equation}
\noindent which correspond to $\tau_{BL5}$ = 6300 and $\tau_{BL10}$ = 93600. These estimations are in agreement with experimental results (Table \ref{resum}).
Note that $\tau$ refer to a global performance estimation (integrated metric of the overall images, see Sect. \ref{results}), and can therefore not inform for a local coronagraphic profile behavior.
Figure \ref{ErrorPlot} gathers experimental and theoretical results for BL10 (s=16) with a broadband filter ($\Delta\lambda/\lambda=24\%$), with simulated profiles obtained assuming independently mask profile error and wavefront error (12 nm rms), as well as a profile including both of these error sources in the same time.
Even if it is difficult to ascertain source of errors, the profile error is consistent with the discrepancy found at small angular separation (i.e. peak level error), while the wavefront errors is probably the dominant source of error in the halo. 
In that situation, the intrinsic limitation of the microdots technique is therefore not a limiting factor.

\section{Conclusion}
We have described the development and laboratory experiment of Band-limited coronagraphs using a microdots design in the near-infrared.
In this paper, we provide design guidelines, and demonstrate the microdots technique as a promising solution for BLC for ground-based observations. 

We have shown with numerical simulations that although total starlight cancellation is not possible, theoretical contrast offer with the microdot approach are deep enough to guarantee that the BLC will not set a limit on the performance
of a ground-based instrument. We note that the theoretical treatment presented in this study does not consider the complexity introduced by potential spectral, and polarization effects of the physical mask.

We identified two sampling configurations suitable for near-IR experiment: $s=16$ and $s=8$, yielding to identical performance in the experiment.
Additionally, we pointed out that the interest of the microdots technique in the light of contrast and IWA requirements is not dependent of the function bandwidth ($\epsilon$) assuming IWA $\geq$ 3$\lambda/D$. This already meets EPICS (20-30 mas in H-band), or SPHERE (0.1 arcsec in H-band) standard requirements. However, we note that stronger requirements of the IWA (1 or 2 $\lambda/D$, ultimate goal of SPHERE), will set a limit on the interest of the technique. 

With prototypes we have demonstrated impressive performance, where limitations have been presumably (on the basis of simulations) introduced either by mask error, or wavefront error of the bench. 
Improvement of the results presented and resumed in Table \ref{resum} are foreseen -- at least for the peak attenuation -- with a new set of prototypes that will provide a better accuracy of the profile in the outer part of the function (calibration issue of the manufacturing process).

Additionally, these raw polychromatic results presented belong as the first tests of BLC in the near-IR, and were managed without active wavefront correction, nor particular data reduction post-processing. Performance reached in the experiment went beyond to the SPHERE prototype performances, APLC results obtained so far with a microdot apodizer on the same bench, and yield to similar contrast than the ones presented in \citet{Creep06} (using $4^{th}$ order notch-filter mask, in a monochromatic visible-wavelength domain) although IWA are not exactly identical.

Ultimately, these final prototypes will be implemented on HOT \citep[the High-Order Testbench, the AO-facility at ESO,][]{HOTbench}, and being compared with others \citep[FQPM, APLC, and Lyots,][]{HOTcorono, microdots1} with atmospheric turbulence generator and AO-correction, as already initiated with intensive simulations \citep{Corono}. Results of this experiment will be presented in a forthcoming paper.

Although this study was carried out in the context of R$\&$D activities for EPICS, it is potentially applicable to upcoming instruments such as SPHERE, or GPI. 
We note that the interest of the technique presented in the paper for space-based operations, is subject to science cases (IWA and contrast requirements). 
For instance, high star-planet contrast ($10^{-10}$ in the visible) at less than 0.1$\arcsec$ (Terrestrial planet), will very likely required a wavefront stability only reachable with a space telescope. In that situation, the technique employed here will impose a limit on the accessible contrast. Notch-filter masks might be more appropriate. 
      
\acknowledgments
The activity outlined in this paper has been partially funded as part of the European Commission, Sixth Framework Programme (FP6), ELT Design Study, Contract No. 011863, and Seventh Framework Programme (FP7), Capacities Specific Programme, Research Infrastructures; specifically the FP7, Preparing for the construction of the European Extremely Large Telescope Grant Agreement, Contract number INFRA-2007-2.2.1.28.

P. M would like to thank Christophe Dupuy from ESO for his helpful support and availability, in particular with metrology inspection, as well as Precision Optical Imaging and Aktiwave (Rochester, New-York) for the delivery of additional prototypes (10-micron specimens). We thank Anthony Boccaletti and the anonymous referee for their useful comments on the present manuscript.
\appendix
\onecolumn
\section{Optical density of a surface layer}
\label{appendix}
A lossy dispersive medium can be described for optical propagation by a complex index  $n + \imath k$ where n and k are related to the linear susceptibility. 
The index n describes the evolution of the phase of the optical wave and k describes the linear attenuation of the optical wave. 
The propagation of an electromagnetic wave of wavelength $\lambda = 2 \pi c$ is described by a complex wavevector $\frac{\omega}{c} \left(n(\lambda) + \imath k(\lambda) \right)$, 
where both the index and attenuation are wavelength-dependent. 

\noindent The input and output electric fields after a propagation distance ($e$) are related by: 
\begin{equation}
E_{out} = E_{in} \times exp\left(\imath \frac{2\pi}{\lambda} (n(\lambda) + \imath k(\lambda)e \right)
\end{equation}

\noindent The intensity transmission ($T = E_{out} / E_{in}$) for a thickness $e$ and wavelength ($\lambda$) is therefore: 
\begin{equation}
T =  exp\left(- \frac{4 \pi k(\lambda) e}{\lambda} \right)
\end{equation}
\noindent The optical density is by definition the base-10 logarithm of the inverse of the intensity transmission, i.e. the natural logarithm of the intensity transmission divided by the natural logarithm of 10: 
\begin{equation}
OD(e, \lambda) = \frac{4\pi k(\lambda) e}{\lambda ln(10)}
\label{OD2}
\end{equation}
\noindent and is therefore dependent on the layer thickness ($e$), the operating wavelength ($\lambda$), and the factor $k$ (material-dependent).

\nocite{*}

\onecolumn
\bibliography{MyBiblio2}

\newpage
\onecolumn

\newpage

\begin{center}
\centering
\begin{figure*}[!ht]
\begin{center}
\includegraphics[width=10.cm]{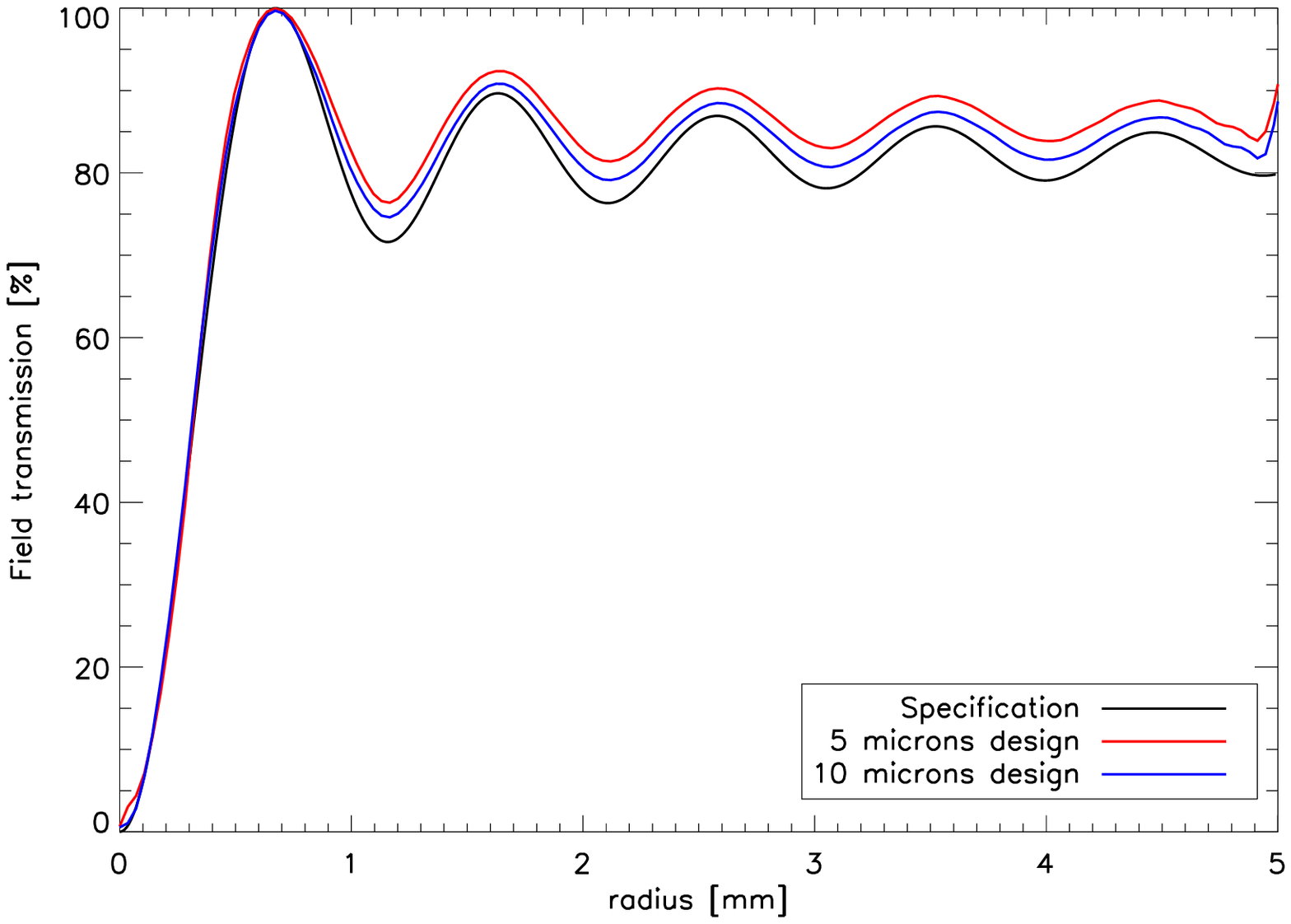}
\includegraphics[width=10.cm]{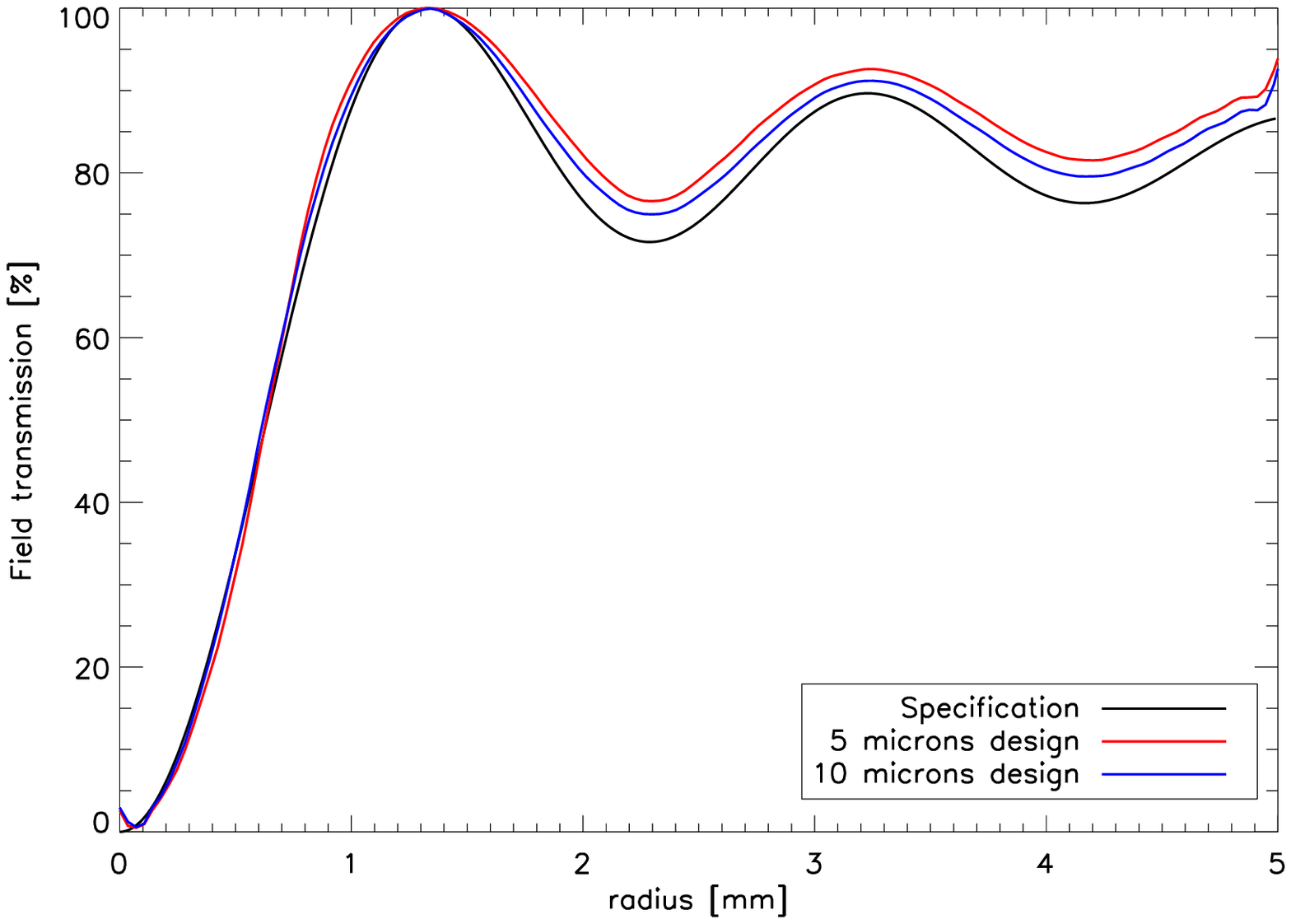}
\end{center}
\caption{Measured profiles of BL5 (top), BL10 (bottom), and specification for the two dot-size prototypes (5 and 10-micron).} 
\label{profile}
\end{figure*}  
\begin{figure*}[!ht]
\begin{center}
\includegraphics[width=5.cm]{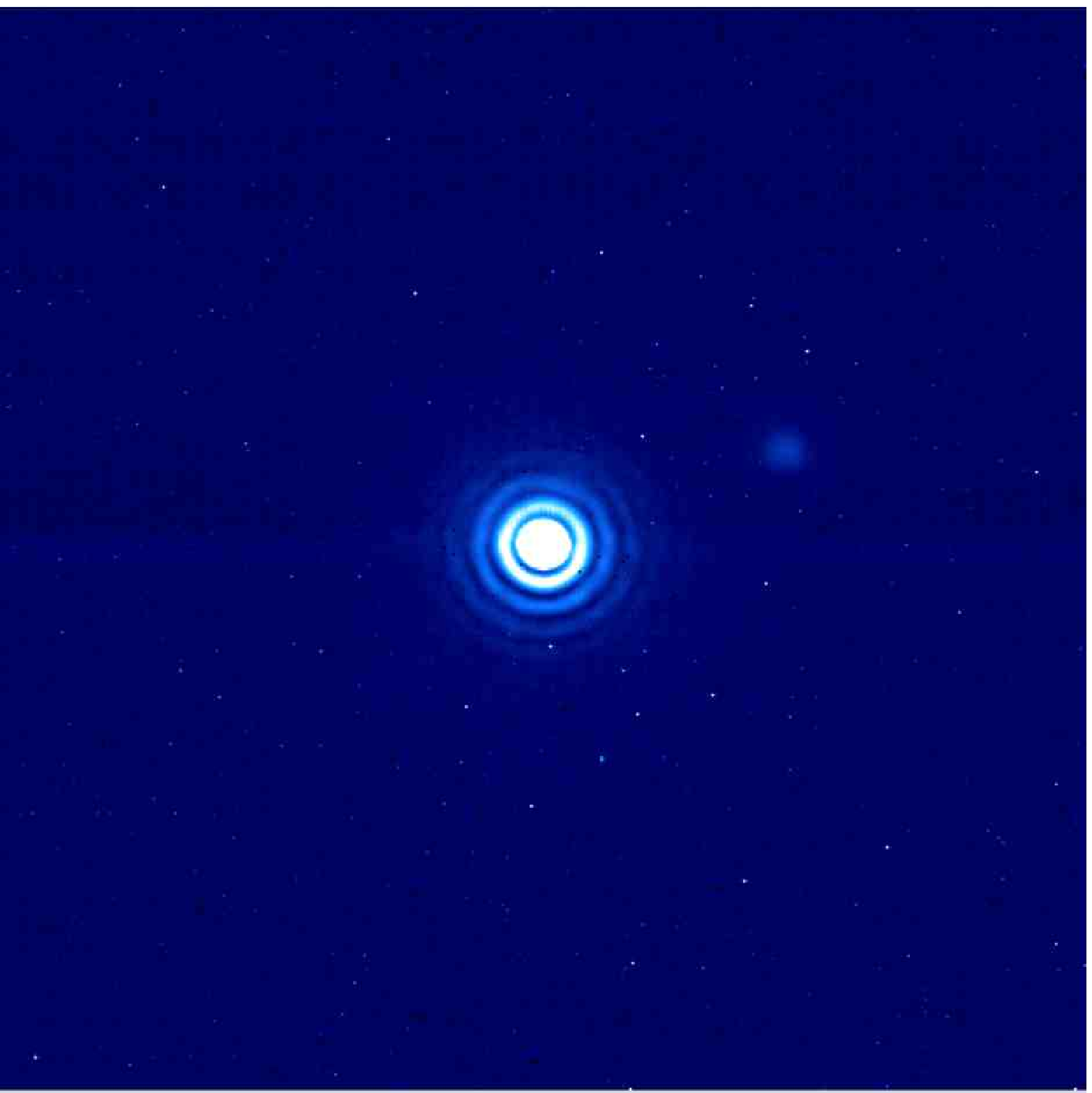}
\includegraphics[width=5.cm]{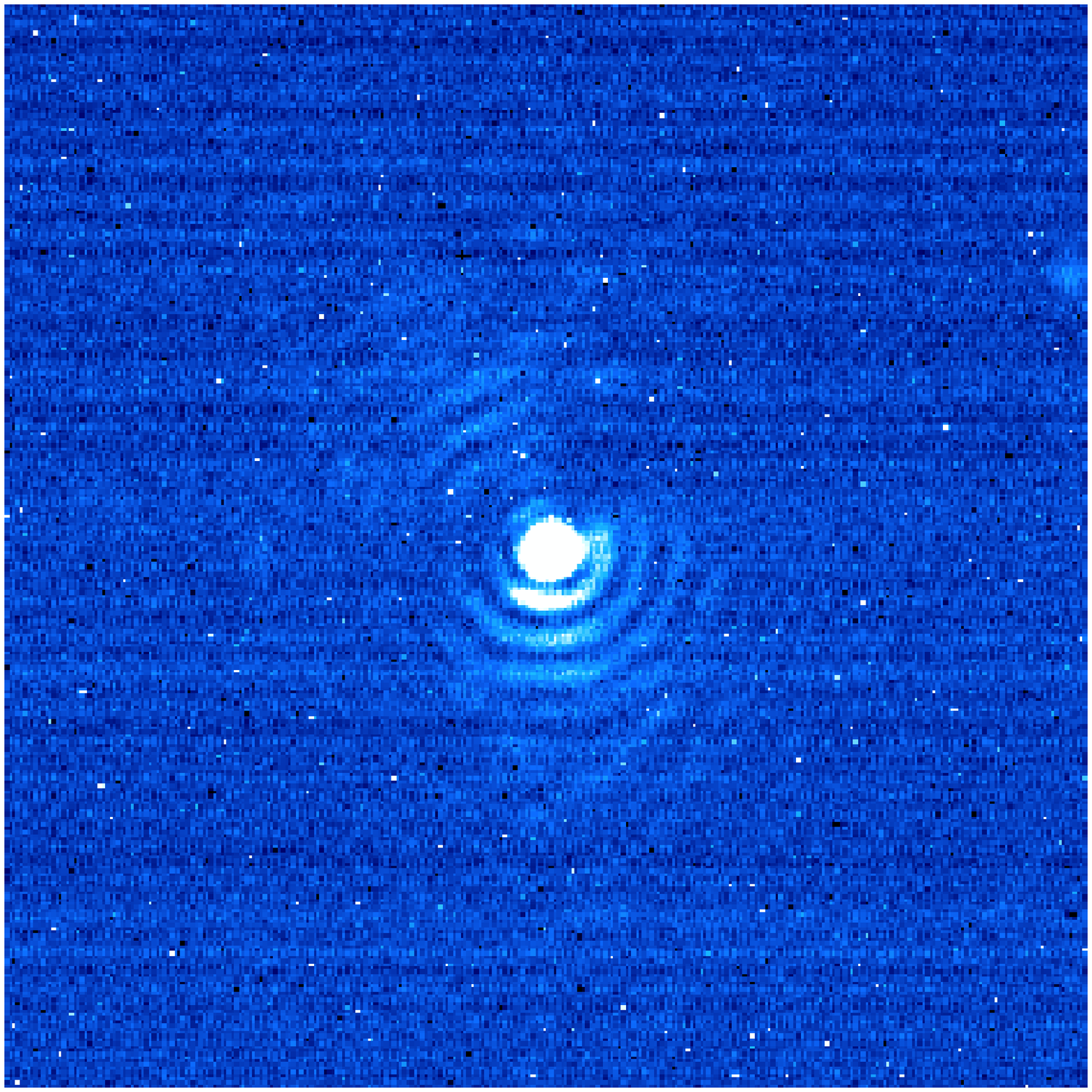}
\includegraphics[width=5.cm]{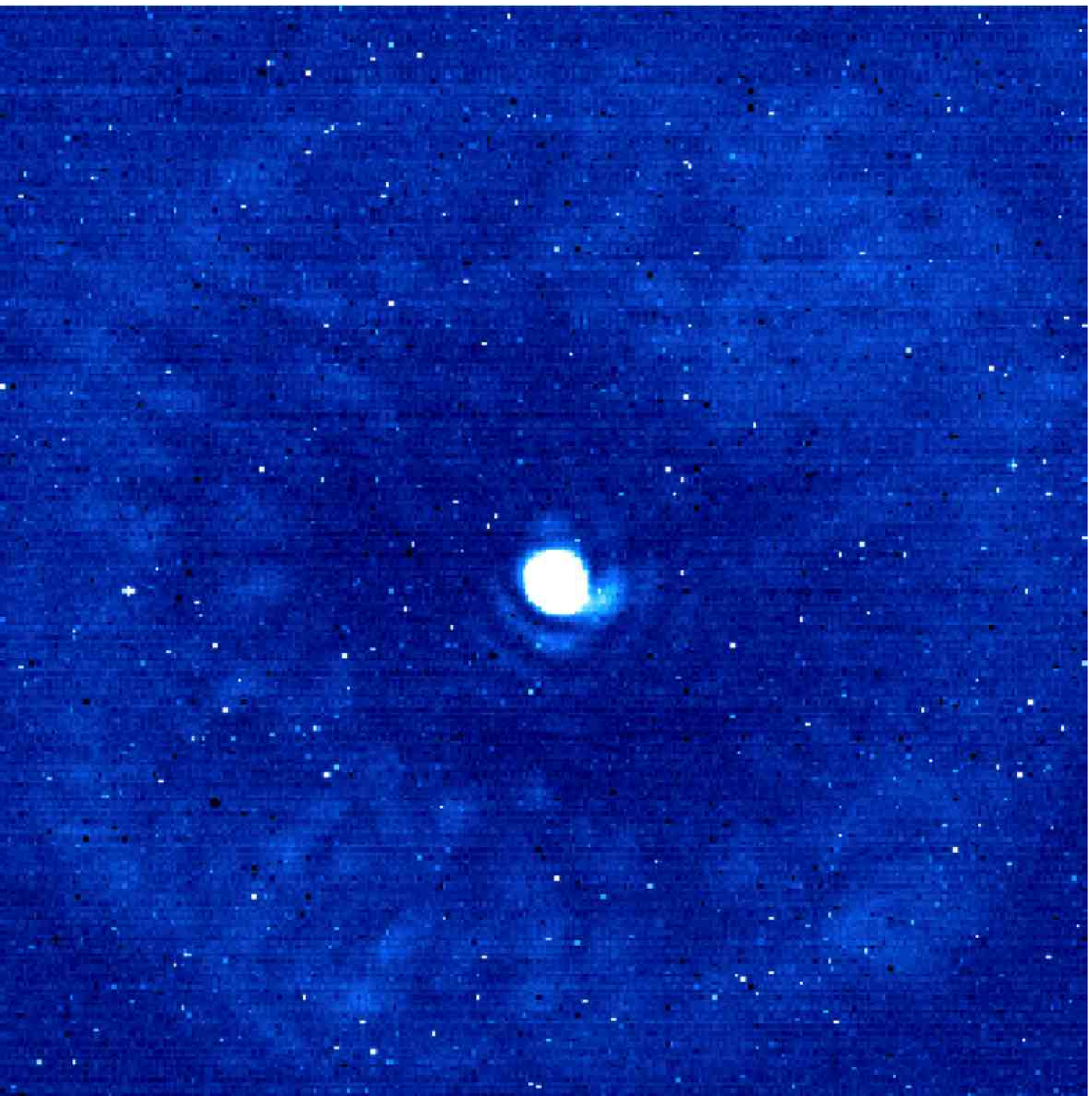}
\end{center}
\caption{From left to right: PSF image (with a ghost, right side of the image), BL5 coronagraphic image, and BL10 coronagraphic image ($\Delta\lambda/\lambda=24\%$). The image dynamic has been chosen to enhance contrast for the sake of clarity.} 
\label{images}
\end{figure*}  
\begin{figure*}[!ht]
\begin{center}
\includegraphics[width=10cm]{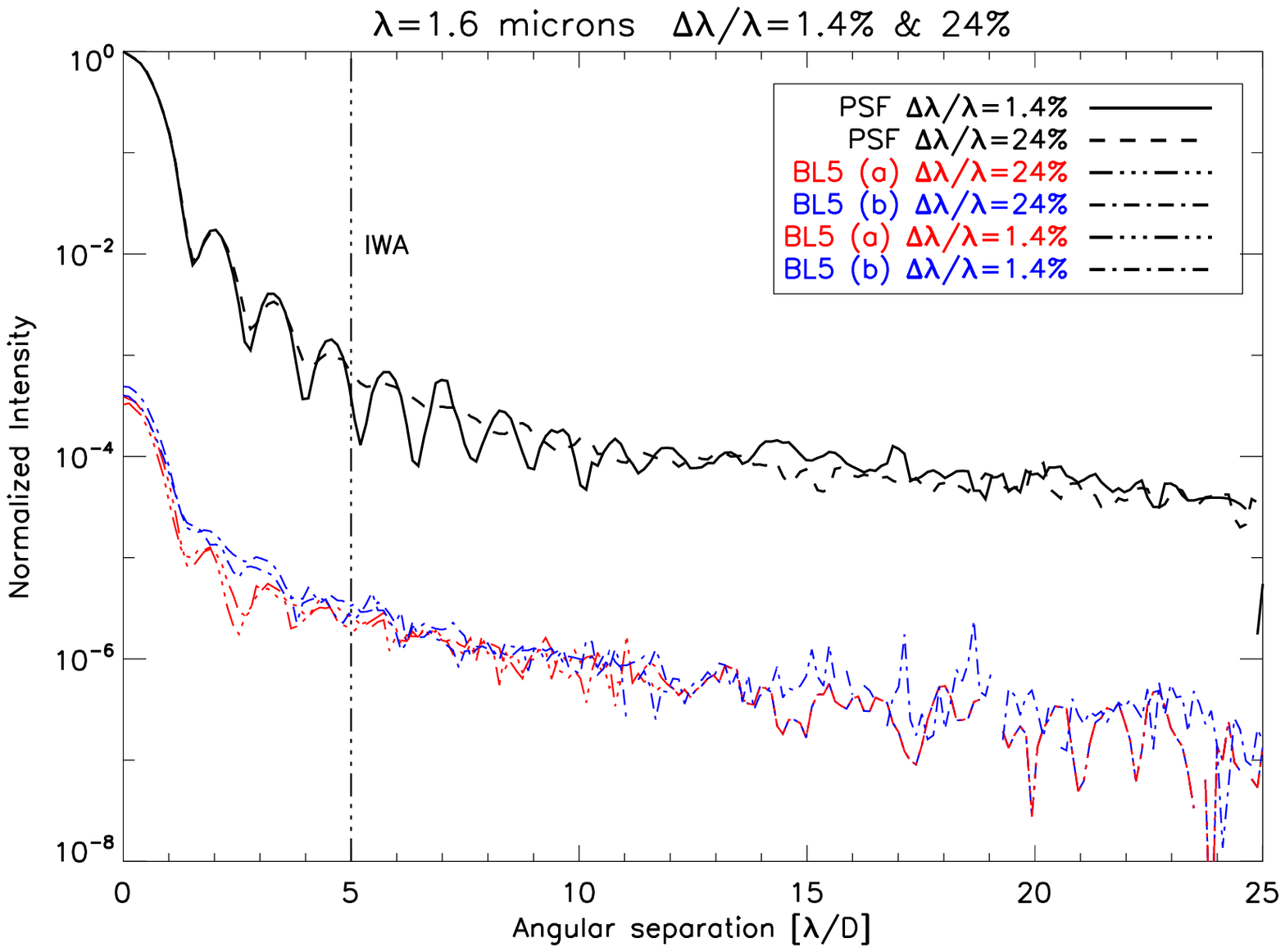}
\includegraphics[width=10cm]{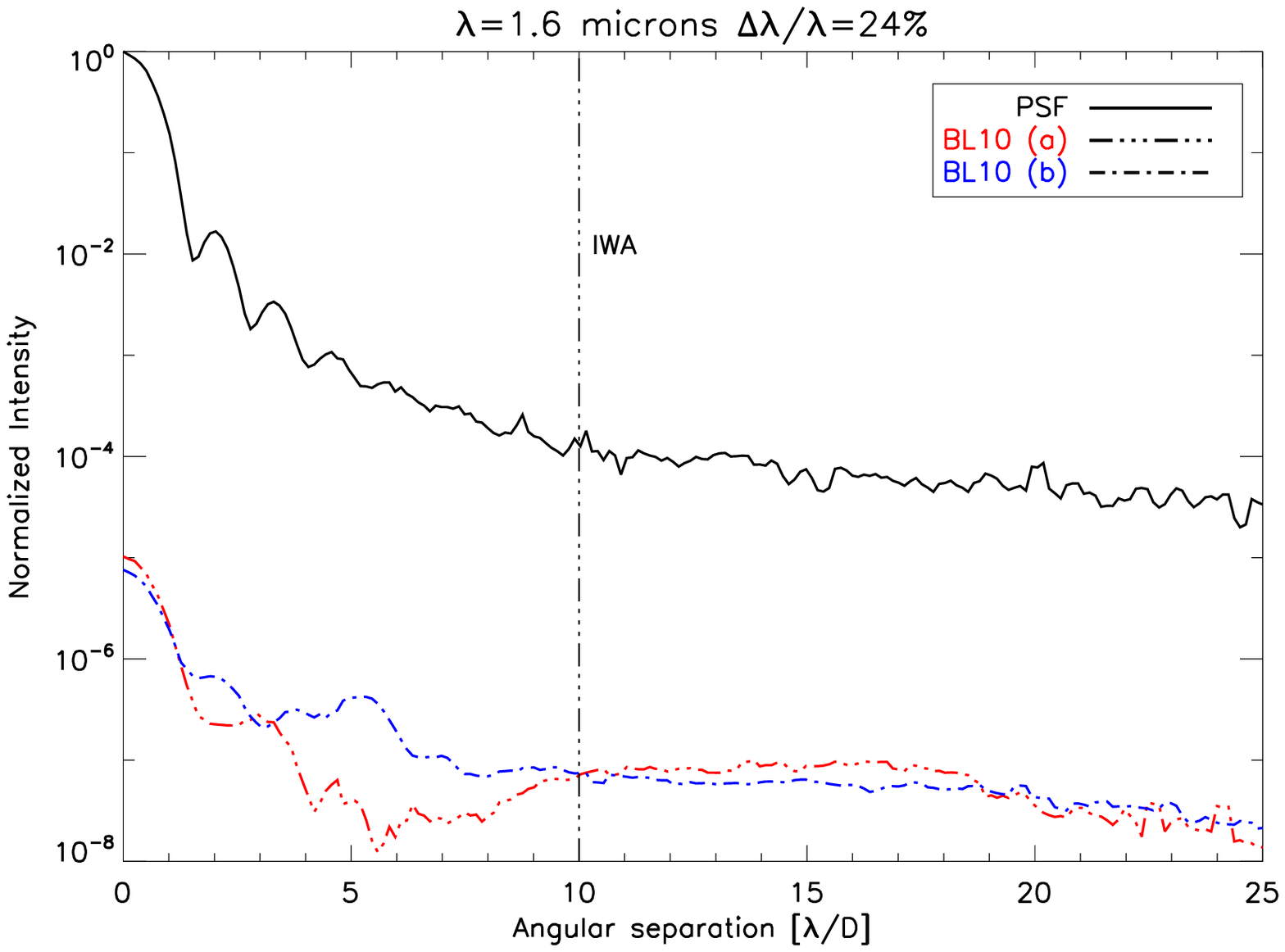}
\end{center}
\caption{Experimental results obtained with BL5 (top) and BL10 (bottom) prototypes, all profiles are azyimuthally averaged.} 
\label{banc}
\end{figure*}  

\begin{figure*}[!ht]
\begin{center}
\includegraphics[width=10cm]{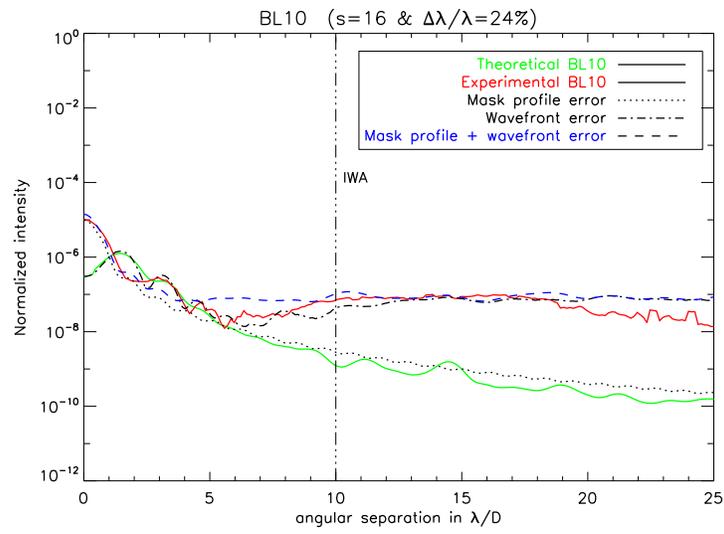}
\end{center}
\caption{Resume of BL10 (s=16 and $\Delta\lambda/\lambda=24\%$) theoretical (green curve) and experimental results (red curve) compared to simulation profiles obtained, when including error sources such as mask profile error (dotted black curve), wavefront error (12 nm rms, full-filled black curve), or both of these errors (blue curve).} 
\label{ErrorPlot}
\end{figure*}  

\newpage
\begin{table}[]
\begin{tabular}{|l|l|l|l|l|l|l|}
\hline 
Metrics & $\tau$ & $\tau_{0}$ & $\mathscr{C}_{IWA}$ & $\mathscr{C}_{12 \lambda/D}$ & $\mathscr{C}_{20 \lambda/D}$ & $\mathscr{C}_{25 \lambda/D}$  \\
\hline
\hline 
BL5 (a) & 2410 & 2554  & $3.7\times10^{-5}$  &    $5.6\times10^{-7}$ & $2.7\times10^{-8}$ & $5.3\times10^{-8}$ \\
BL5 (b) & 1804 & 2038 & $3.0\times10^{-5}$  &    $8.1\times10^{-7}$ & $2.6\times10^{-7}$ & $8.5\times10^{-8}$ \\
\hline
\hline
BL10 (a) &  81606 & 97375  & $1.5\times10^{-7}$  &  $7.7\times10^{-8}$ & $3.7\times10^{-8}$ & $1.3\times10^{-8}$ \\
BL10 (b) & 83177 & 131809 &  $1.3\times10^{-7}$  & $6.3\times10^{-8}$ & $4.3\times10^{-8}$ & $2.1\times10^{-8}$ \\
\hline
\end{tabular}
\caption{Summary of coronagraphic results obtained with BL5 and BL10 ($\Delta\lambda/\lambda=24\%$).}
\label{resum}
\end{table}
\end{center}

\end{document}